\documentstyle[twocolumn,prl,aps]{revtex}
\begin{document}
\draft
\title{
Chen and Itoi Reply
}
\author{Wei Chen}
\address{Department of Physics,
University of North Carolina,
Chapel Hill, NC 27599-3255}
\author{Chigak Itoi}
\address{Department of Physics and Atomic Energy Research Institute,
College of Science and Technology,\\ Nihon University,
Kanda Surugadai, Chiyoda-ku, Tokyo 101, Japan
}
\maketitle

\begin{narrowtext}
In a Comment \cite{H} statements on
our paper \cite{CI} are made with which we disagree.
\end{narrowtext}

It is argued in \cite{H} that the model (3) can not
be equivalent to (1) and (2), as
the former has twice as many degrees of freedom
as the latter.
To clarify the issue, let us recall two key ingredients
to claim the equivalence in \cite{CI}
and a sequent work \cite{CI1}:
I) introducing the coherent states and integrating out the
matter and CS fields, we obtain a
continuum random walk expression for
the partition function, {\it e.g.} eq. (9) in \cite{CI},
in which the spin of matter field
$j$ appears as the coefficient of spin factor
and the CS coefficient $\alpha$ as the coefficient
of relative and self-energies;
and II) there have been topological relations between
the spin factor
and relative and self-energies regularized
by point splitting, eqs. (12) and (13).
The expression and topological relations
suggest that an integer (or odd-half-integer)
part of $\alpha$, $n$, can be absorbed by changing $j$.
Then, as the spin
$j$ characterizes the degrees of freedom carried
by the matter field, the latter
may be changed when the spin-$j$
matter is mapped to spin-$(j+n)$ matter
and $\alpha$ is shifted
to $(\alpha - n)$. This
indicates that
CS interactions
affect the degrees of freedom
of the matter field and whole system.
After all, the equivalent CS models
describe the same (anyon) system and
have the same amount degrees of freedom.
While it has been known in many (1+1)
dimensional field theories that interactions
change the degrees of freedom carried by fields,
it seems for us to have seen a possibility for
such an example in (2+1) dimensions.

It is pointless to require equivalent models to involve
same divergences. However, as a
matter of fact, the models (1), (2) and (3)
all need use of ultraviolet regularization,
including in calculating $\Pi_o$
in (1), contrary to that argued in \cite{H}.
The author of \cite{H} further picks up
difference between one-loop
corrections to $\Pi_o(p)$
calculated in a particular regularization
 in (1) and (2)
to argue inequivalence. We have two remarks to make.
First, it makes no sense
to compare non-universal quantities
such as $\Pi_o(0)$'s in different models. Since these are
regularization dependent, they can take any values that
symmetries allow.
Though, it is tempted to check the local
relevant operator  content. Non-vanishing
$\Pi_o(0)$'s and $\Pi_e(0)$'s
imply that the current-current
correlations in the low energy limit
induce a  CS term and a Maxwell term in
both (1) and (2).
Secondly, the non-local terms
from $\Pi_o(p\neq 0)$ and
$\Pi_e(p\neq 0)$ are
finite in both (1) and (2),
as they should be.
Nevertheless, one should be aware that for any given $\alpha$, at
least two of the three models are  strongly coupled ones, and thus
it is difficult to check non-perturbative results
by means of perturbation, even qualitatively sometimes.
In particular, it makes no judge to the equivalence
to simply compare the values of those non-local
terms obtained in perturbation at one loop.

To argue the models (1) and (2) dissimilar,
the author of \cite{H}
takes $\alpha \rightarrow \infty$
and discusses some process, 
as if there existed no CS terms.
It is inconsistent at all, as large $\alpha$'s
amount to strong CS couplings
that can not be ignored.
This is readily seen by
rescaling
$a_\mu \rightarrow \sqrt{4\pi\alpha}a_\mu$
for (1), for
instance, and the CS fermion coupling has
the strength $\sqrt{4\pi\alpha}$.
Obviously, when $\alpha \rightarrow \infty$,
perturbation around the free fermion theory
is meaningless.
An effective approach to handle strong
CS interactions
has been suggested in \cite{CI} and \cite{CI1}.
That is to trade the integer part of
$\alpha$ for higher spin of matter fields, and then
perturbative calculations in the remaining weak
CS interaction could be good approximations.
In this case, taking $\alpha \rightarrow \infty$,
the models (1), (2) and (3) are now
all describing
a spin-$\infty$ field coupled
to the external field $C_\mu$,
and are thus formally equivalent
(though it is not clear to us
whether the limit is well-defined).
On the other hand, if taking weak
coupling limit in one model,
$\alpha\rightarrow 0$ for instance,
the CS field is decoupled  in (1), while
CS interactions are considerably strong in
(2) and (3).

In conclusion, the author of \cite{H} seems to have overlooked the
highly non-trivial CS couplings, and his hand-waving arguments
are totally groundless.

The work of W.C. was supported in part by
the  U.S. DOE grant No. DE-FG05-85ER-40219.

\end{document}